\documentclass[aps,prl,reprint]{revtex4-2}
\usepackage{graphicx}
\usepackage{dcolumn}
\usepackage{bm}
\usepackage{float}
\usepackage{amssymb}
\usepackage{xcolor,soul} 
\usepackage{braket}
\usepackage{hyperref}
\usepackage[utf8]{inputenc}
\usepackage[T1]{fontenc}

\newcommand\Ca{$^{40} \text{Ca}^+$}

\draft 

\begin{document}

\title{Individual-Ion Addressing and Readout in a Penning Trap}

\author{Brian J. McMahon}
\email{brian.mcmahon@gtri.gatech.edu}

\author{Kenton R. Brown}
\author{Creston D. Herold}
\author{Brian C. Sawyer}%
\affiliation{Georgia Tech Research Institute, Atlanta, GA 30332, USA}%

\date{\today}
\begin{abstract}
We implement individual addressing and readout of ions in a rigidly rotating planar crystal in a compact, permanent magnet Penning trap. The crystal of \Ca is trapped and stabilized without defects via a rotating triangular potential. The trapped ion fluorescence is detected in the rotating frame for parallel readout. The qubit is encoded in the metastable D$_{5/2}$ manifold enabling the use of high-power near-infrared laser systems for qubit operations. Addressed $\sigma_z$ operations are realized with a focused AC Stark shifting laser beam. We demonstrate addressing of ions near the center of the crystal and at large radii. Simulations show that the current addressing operation fidelity is limited to $\sim 97\%$ by the ion's thermal extent for the in-plane modes near the Doppler limit, but this could be improved to infidelities $<10^{-3}$ with sub-Doppler cooling. The techniques demonstrated in this paper complete the set of operations for quantum simulation with the platform.
\end{abstract}

\maketitle 

Quantum simulation of many-body spin models has undergone rapid development in the noisy intermediate scale quantum (NISQ) era. Demonstrated quantum simulation hardware platforms include neutral atoms in optical tweezers \cite{ebadi_quantum_2021,graham_multi-qubit_2022}, RF Paul and Penning trapped ions \cite{monroe_programmable_2021}, and superconducting qubits \cite{kjaergaard_superconducting_2020,harrigan_quantum_2021}. The platforms vary in their accessible qubit number, primitive operation fidelity, and native qubit connectivity. Penning traps and neutral atom arrays have both been employed in experiments with hundreds of qubits. Due to their large trap depths, ion traps exhibit long confinement times, and lifetimes longer than weeks have been demonstrated in a Penning trap \cite{mcmahon_doppler-cooled_2020}. Ion crystals permit full connectivity through collective motional modes \cite{britton_engineered_2012,zhang_observation_2017,joshi_exploring_2023}, while other platforms construct connectivity via local interactions \cite{harrigan_quantum_2021}. 

As discussed in Refs. \cite{blatt_quantum_2012,kranzl_controlling_2022,bruzewicz_trapped-ion_2019,kiesenhofer_controlling_2023}, the lack of individual-qubit addressing in Penning trap systems, due to continuous crystal rotation, has thus far constrained the range of accessible spin-spin couplings. In this Letter, we demonstrate single-ion addressing operations and individual readout of metastable qubit states in planar, defect-free trapped-ion crystals confined in a compact permanent magnet Penning trap. 

\begin{figure}
    \centering
    \includegraphics[width=\linewidth]{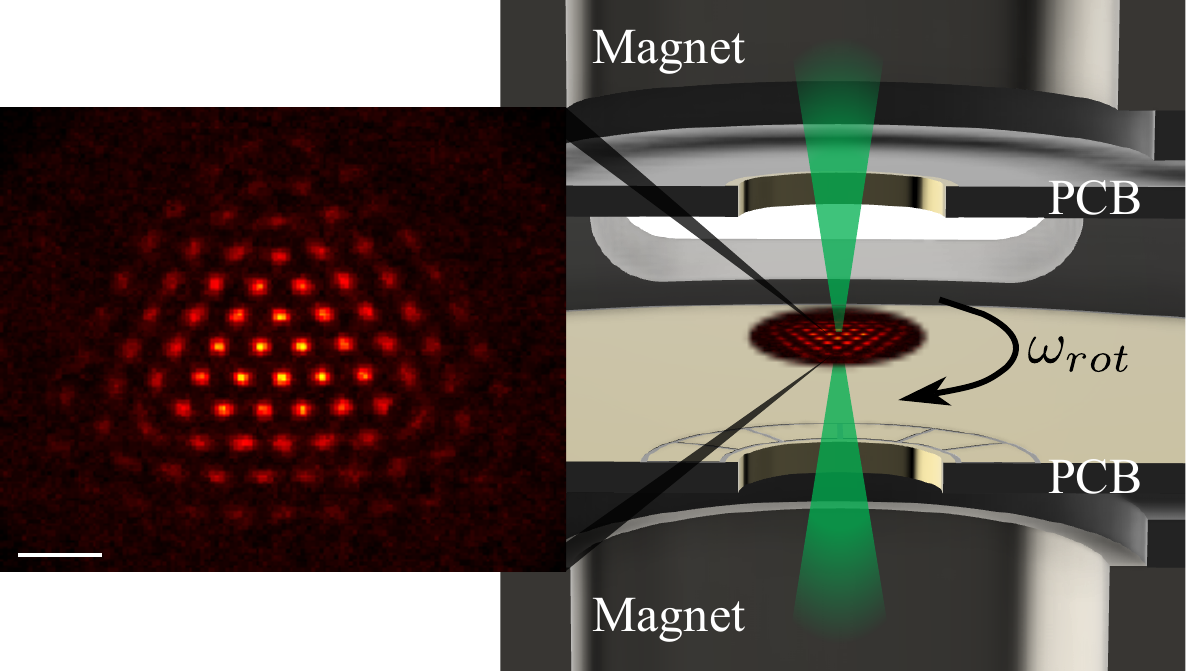}
    \caption{(left) A derotated, top view camera image (0.5 s exposure) of a trapped \Ca crystal of $\sim 93$ ions rotating at $50$ kHz. The scale bar corresponds to $50\ \mu m$. (right) Cross-section illustration showing the permanent magnets, the trap electrodes (on printed circuit boards), an example ion crystal, and an addressing beam intersecting the crystal plane. Note that the ions in the model are not to scale.}
    
    \label{fig:tricrystal}
\end{figure}

The ions in a Penning trap are confined in the radial direction by the Lorentz force. Thus, the ions continuously rotate about the trapping magnetic-field lines. The ions are confined along the magnetic field axis (axial direction) by an electric quadrupole potential. For the work described here, a pair of permanent magnets generates a $0.91$~T magnetic field. Gold electrodes on a pair of printed circuit boards (PCBs) shown in Fig.~\ref{fig:tricrystal} provide a confining electrostatic quadrupole potential. The electromagnetic fields of our compact trap produce single-\Ca~motional frequencies of order $20$ kHz, $120$ kHz, and $330$ kHz for the magnetron, axial, and modified cyclotron modes, respectively. When cooled to sufficiently low temperatures, a trapped ensemble of ions forms a rotating Coulomb crystal whose conformation can be directly controlled via applied torques. Demonstrations of this control have used a radial cooling beam offset from the trap center \cite{itano_perpendicular_1988,mavadia_control_2013}, a ``rotating wall'' potential \cite{torrisi_perpendicular_2016}, or both. A planar crystal is formed for a range of crystal rotation frequencies where axial confinement is much stronger than radial confinement \cite{wang_phonon-mediated_2013}. These crystals are typically controlled using a quadrupolar rotating wall potential, which enables phase-synchronous rotational control of the array \cite{hasegawa_stability_2005}. However, a quadrupolar boundary condition does not match the crystal conformation's lowest energy configuration: a triangular lattice. Due to the finite ion temperature, the induced crystal dislocations reconfigure on experimental timescales \cite{dubin_structure_2013}. Thus, a rotating triangular potential was proposed to stabilize the ion array \cite{khan_theoretical_2015,dubin_structure_2013}. Here, we demonstrate this technique using 12-way azimuthally, segmented electrodes that are patterned onto the PCBs which face the ions. Ions are confined between the two PCBs, which are spaced by $5$~mm. They each have a $4.24$~mm diameter central hole for optical access. The segmented electrodes on the PCBs are shown in the section view of the schematic in Fig. \ref{fig:tricrystal}. The ion positions are found to be well-localized in the rotating frame; no dislocations in the crystal are observed (e.g. camera image of Fig. \ref{fig:tricrystal}), confirming the efficacy of the rotating triangular potential.

Ions in a given planar crystal are detected in their rotating frame such that they appear stationary (``derotated"). A photon intensifier, located between the ions and the camera, multiplies the incoming, $397$~nm photons from the trapped ions. A camera \footnote{Tpx3Cam from Amsterdam Scientific Instruments with a photon intensifier from Photonis} records the positions and arrival times of these signals (hits). A custom analysis server collects data from the camera and performs a coordinate transformation into the rotating frame. The analysis server has a calibrated set of predefined regions-of-interest for each ion. For each fluorescence detection interval, the server counts the number of hits in each region. Alternatively, one can retrieve a composite image of all hits after derotation is performed, as shown in Fig. \ref{fig:tricrystal}. Every experiment is triggered at the same phase of the crystal's rotation. Thus, at any given time in the experiment, the crystal orientation is well defined. Similar derotation of Penning trap crystals was demonstrated in \cite{wolf_efficient_2023,bohnet_quantum_2016,britton_engineered_2012}.

\begin{figure}
    \centering
    \includegraphics[width=0.75\linewidth]{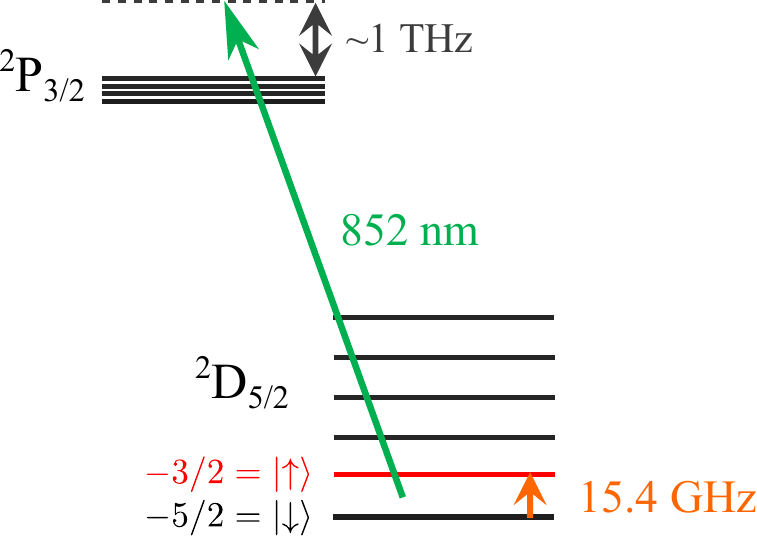}
    \caption{The \Ca~levels used for implementation of a metastable qubit. An 852 nm laser creates a differential AC Stark shift of the qubit levels. Transitions between the levels are driven directly with microwave radiation near $15.4$ GHz.  }
    \label{fig:leveldiagram}
\end{figure}

For this work, two metastable Zeeman states in the D$_{5/2}$ manifold comprise the qubit states, as shown in Fig. \ref{fig:leveldiagram}. These states are spectrally well-resolved by the trapping magnetic field of $0.91447$ T. For each experiment, the m$_J=-5/2$ ($\ket{\downarrow}$) state is prepared through optical pumping using electric-dipole transitions near $393$ nm, $397$ nm, $854$ nm, and $866$ nm. A rate equation simulation of this state preparation technique predicts a fidelity of $99.5\%$. Global one-qubit operations are driven between the two states using microwave radiation near $15.4$ GHz delivered through a microwave horn. The second-order Zeeman shift lifts the degeneracy of the Zeeman splittings between neighboring transitions in the D$_{5/2}$ manifold by $\sim 14$ MHz, spectrally isolating the qubit transition. To read out the qubit state, the m$_J=-3/2$ ($\ket{\uparrow}$) population is optically pumped to the S$_{1/2}$ manifold with a typical fidelity of about $95\%$, as confirmed by numerical simulations \footnote{This can be improved by deshelving the population directly using a $729$ nm laser resonant with S$_{1/2}$ to D$_{5/2}$ electric quadrupole transition}.

A recently-proposed addressing method for ions in Penning traps relies on distortions of global laser beams \cite{polloreno_individual_2022}. In this work, addressing of individual ions in the crystal is performed instead using a focused laser beam with a waist smaller than the inter-ion spacing. Compared to traditional superconducting magnets, the reduced form factor of our compact Penning trap permits straightforward implementation of optical ion addressing protocols \cite{mcmahon_doppler-cooled_2020}. For the experiments in this Letter, the addressing beam has a $<25\ \mu$m waist ($1/e^2$ intensity radius) which produces local AC Stark shifts (ACSS) on individual qubits. The addressing beam is derived from an $852$ nm laser stabilized to the D2 transition in neutral cesium. The Cs D2 transition frequency is conveniently detuned from the resonant D$_{5/2}$ to P$_{3/2}$ transition by about $+1$ THz.

\begin{figure}
    \centering
    \includegraphics[width=\linewidth]{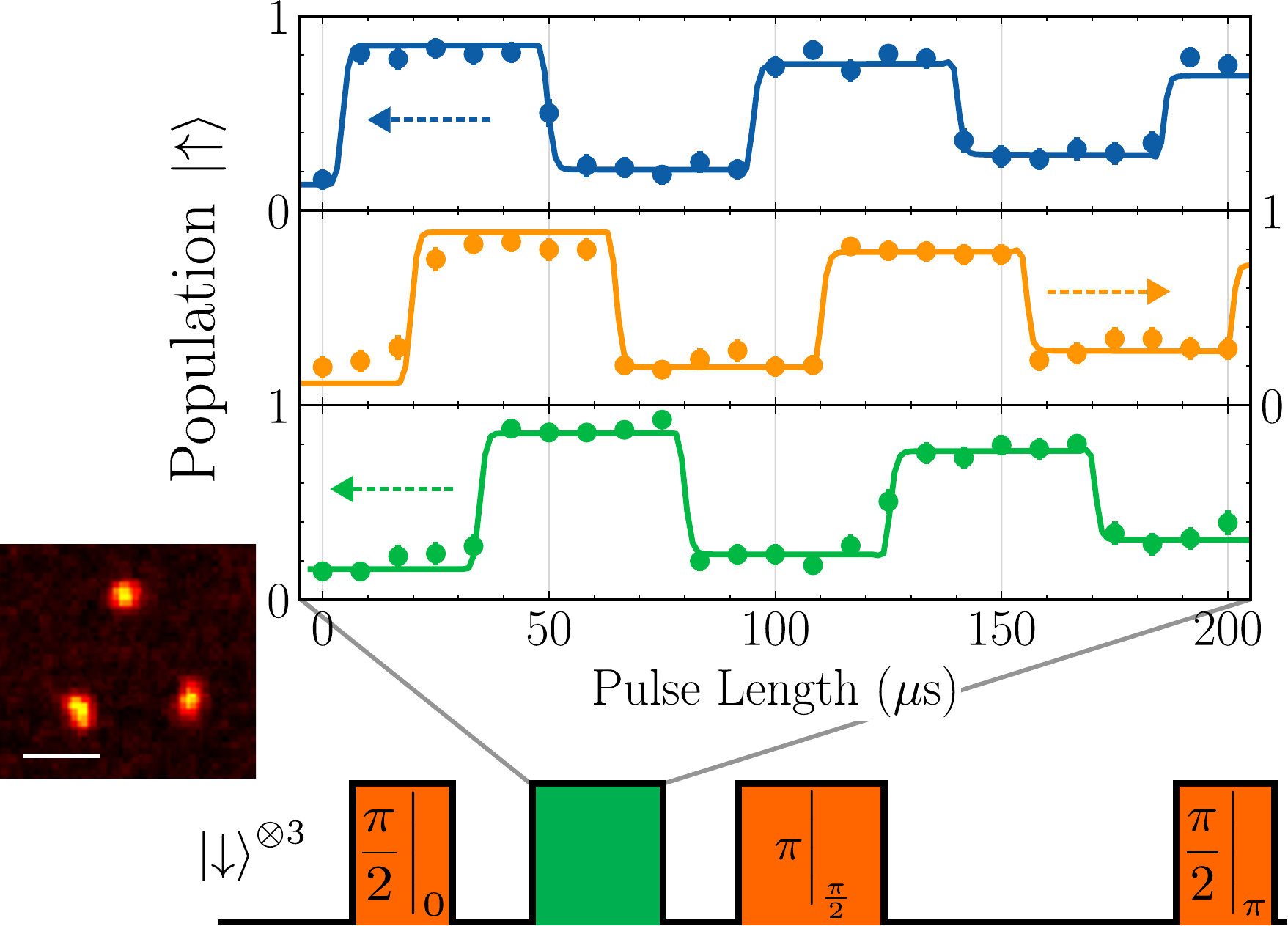}
    \caption{Individual ion populations at the end of a Ramsey sequence as the addressing pulse duration is varied. The pulse is applied within the first arm of a Ramsey spin echo sequence. The simulation (solid lines) uses measured experimental parameters, a wall amplitude of $1.5$ V, and an average in-plane mode temperature of $0.63$ mK. The orange squares show the global microwave pulses, which rotate all qubits by an angle of $\pi/2$ or $\pi$ on the Bloch sphere. The green block represents the addressing ACSS pulse. The inset shows a derotated camera image of the crystal used for this data with a $50\ \mu m$ scale bar.}
    \label{fig:cwaddressing}
\end{figure}

We characterize the single-ion addressing pulses using a global Ramsey spin-echo sequence acting on a three-ion crystal, where the addressing pulse is applied only during one arm of the sequence (Fig. \ref{fig:cwaddressing} bottom). For short pulse durations, none of the ions are illuminated by the addressing beam. For longer pulse durations, the crystal rotation ($22$ kHz, $45.5\ \mu$s period) brings each ion successively through the beam at $15\ \mu$s intervals. The beam intensity is calibrated to induce a phase shift of $\pi$ during a single $\approx 5\ \mu$s beam transit. Each $\pi$ phase shift manifests as an inversion of the population at the completion of the Ramsey sequence. The addressing beam is focused at the measured ion radius of $48(1)\ \mu$m and has an asymmetric beam waist of $23(2) \times 13(2)\ \mu$m, with the narrower dimension predominantly along the ion trajectory. The data of Fig. \ref{fig:cwaddressing} show the population in $\ket{\uparrow}$ of each of three ions after this sequence, where regular inversions are visible at the expected intervals. The figure also shows an image of the crystal used. 

The solid lines on the graph in Fig. \ref{fig:cwaddressing} show simulations of the final $\ket{\uparrow}$ populations accounting for the measured SPAM errors, decoherence without addressing, and thermal in-plane ion motion. The simulation assumes a three-ion crystal with $116$ kHz axial frequency, a wall amplitude of $1.5$ V, an in-plane mode temperature of $0.63$ mK (near the Doppler limit), and an AC Stark shift of $200$ kHz at the beam center \footnote{For this intensity and detuning, off-resonant photon scatter predominantly causes qubit leakage to the S-manifold at $\sim 80\ s^{-1}$ during each $5\ \mu$s addressing transit.}. The simulations begin with a numerical calculation of the equilibrium ion configuration for a crystal in the rotating frame, based on the models detailed in Refs. \cite{shankar_broadening_2020,dubin_normal_2020}. The simulated equilibrium positions of the three ions lie on a circle of radius $\sim 47.1\ \mu$m, which agrees with the independent measure of the ion radius using the addressing beam ACSS interaction. Using the numerical solution to the crystal configuration, the in-plane mode frequencies and mean-square thermal displacements are calculated accounting for the negative energy contribution of the magnetron mode and the positive energy of the cyclotron modes. In simulations of the gate interaction, an ion is rotated through the addressing beam at the crystal's rotation frequency. An integral is evaluated to calculate the phase accumulation as the ion transits the addressing beam for a given laser beam intensity, frequency, and waist. At zero temperature, the laser beam intensity and position may be calibrated for repeatable, high-fidelity $\pi$ rotations about the z-axis on the Bloch sphere. In order to simulate the effects of finite temperatures on the addressing interaction, the ion's position includes a sum over the oscillations along each mode direction with the appropriate frequency, random phase, and mean thermal displacement. For each pulse duration, we average $100$ randomized simulation repetitions.

\begin{figure}
    \centering
    \includegraphics[width=\linewidth]{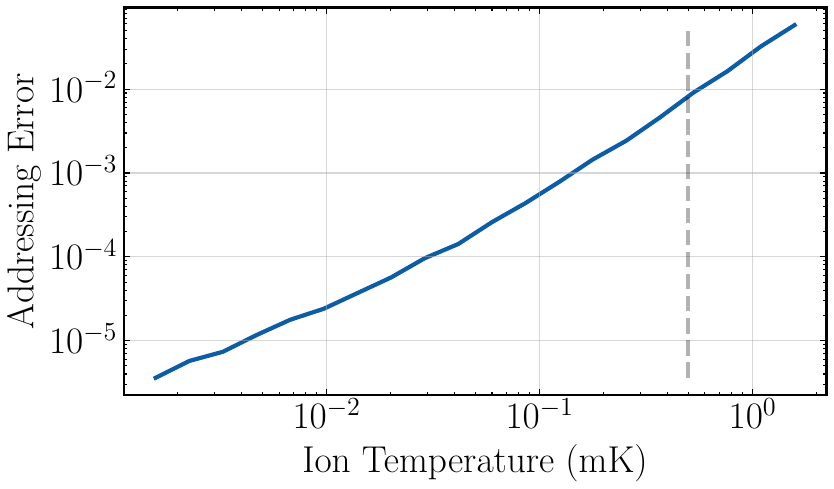}
    \caption{For the crystal configuration shown in Fig. \ref{fig:cwaddressing}, the $\pi$ pulse infidelity is simulated as a function of the temperature for each mode. The vertical dashed line shows the Doppler limit for the cooling transition in \Ca.}
    \label{fig:addrfidelity}
\end{figure}

The simulations show that the in-plane ion temperature plays a large role in the fidelity of the addressing operations. Using the same ion configuration parameters for the crystal shown in Fig. \ref{fig:cwaddressing}, the addressing error can be simulated for crystals with varied temperature. At the Doppler limit of $\sim 1$ mK \cite{mavadia_optical_2014}, the average thermal occupation implies average thermal extents of $<300$ nm for the modified cyclotron branch of modes ($310-309$ kHz) and $<6\ \mu$m for the magnetron modes ($2.5-1.1$ kHz). The thermal extent of the lowest frequency mode, (i.e. `rocking mode'), scales in frequency with the rotating wall potential \cite{tang_equilibration_2021}. This mode's extent can be $>10\ \mu$m for wall strengths that are a small fraction of the trapping voltage. When the temperatures of the lower frequency modes result in mean thermal displacements that are a large fraction of the addressing beam size, then the addressing ACSS varies across experiment realizations. This reduces the average fidelity of a single $\pi$ transition. Higher mode frequencies or sub-Doppler cooling of the in-plane modes could circumvent this issue \cite{johnson_rapid_2024}.
Simulations shown in Fig. \ref{fig:addrfidelity} suggest that the residual in-plane mode oscillation amplitudes would have a contribution to the infidelity $<10^{-5}$ as the ion mode temperatures $< 40\ \mu$K.

\begin{figure}
    \centering
    \includegraphics[width=\linewidth]{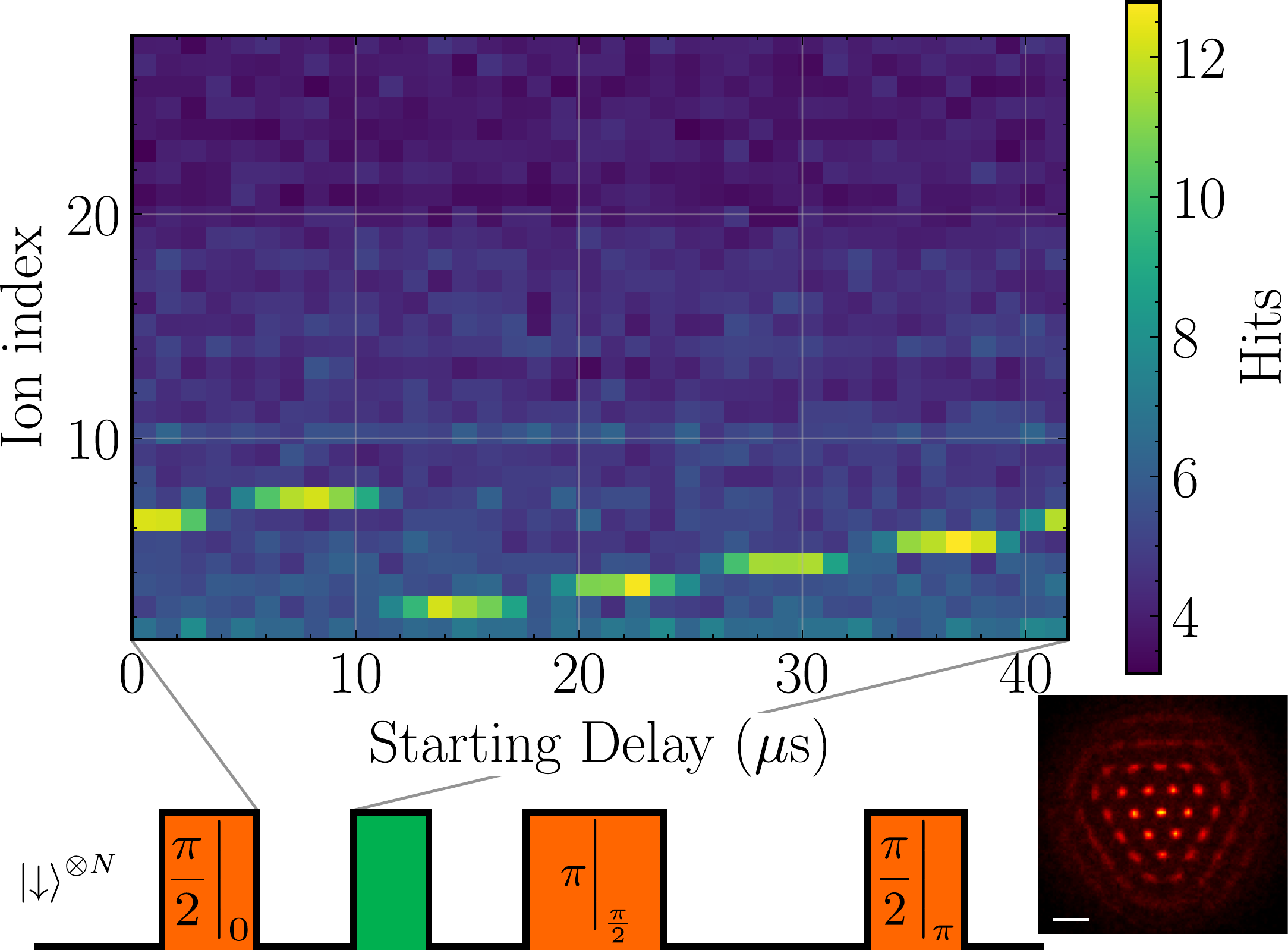}
    \caption{Measured hit numbers for 27 ions at the end of a Ramsey sequence as the starting delay is varied before a $5\ \mu$s addressing pulse. The addressing pulse is applied within the first arm of a Ramsey spin echo sequence. The addressing laser beam is focused at the shared radius of the first six ions. The orange squares show the global microwave pulses, which rotate all qubits by an angle of $\pi/2$ or $\pi$ on the Bloch sphere. The green block represents the addressing ACSS pulse. The inset shows a derotated camera image of the crystal used for this data with a $50\ \mu m$ scale bar.}
    \label{fig:pulsedaddressing}
\end{figure}

\begin{figure}
    \centering
    \includegraphics[width=\linewidth]{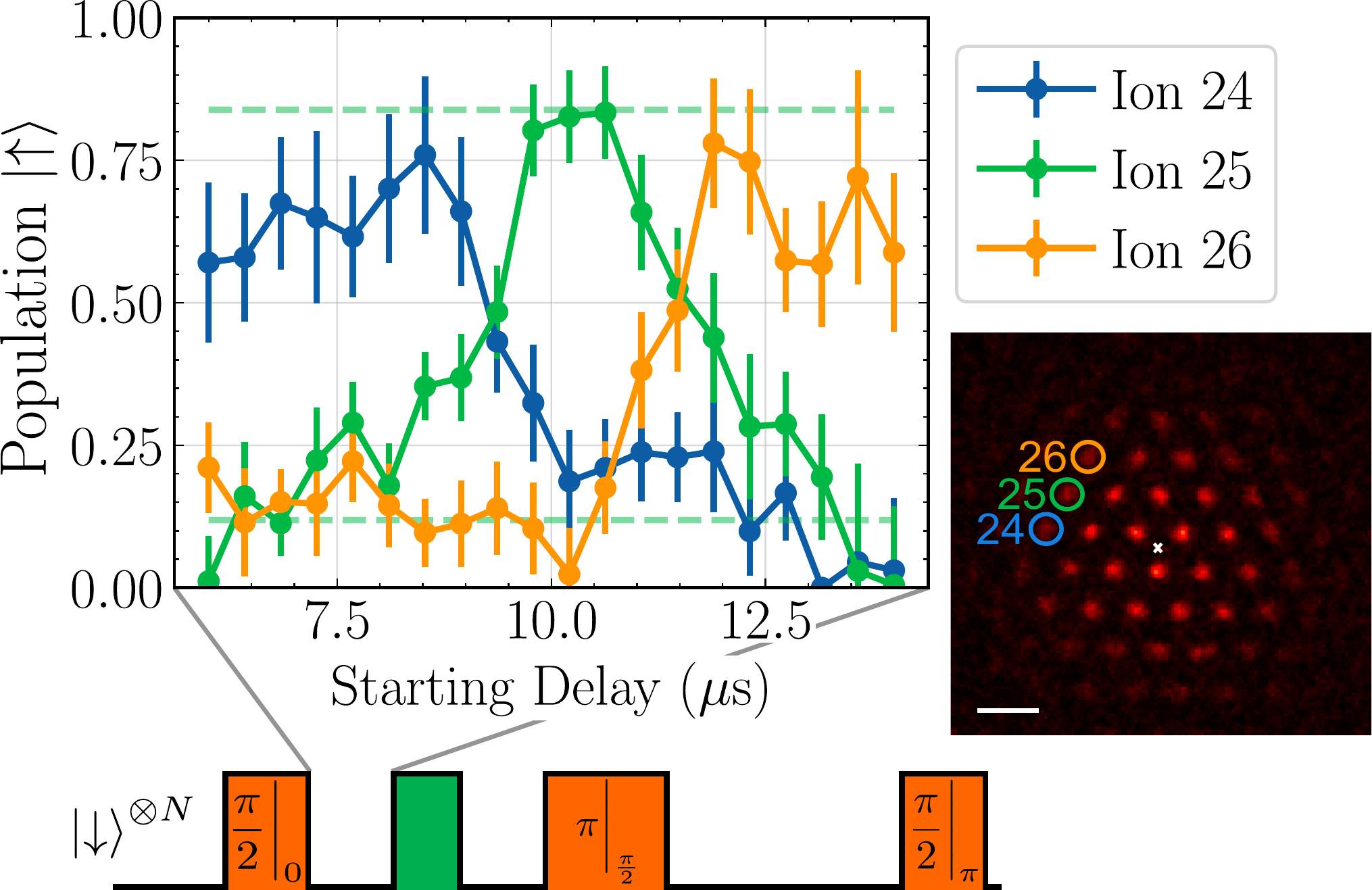}
    \caption{Individual ion populations for ions 24, 25, and 26 at the end of a Ramsey sequence as the addressing pulse starting delay is varied. The pulse is applied within the first arm of a Ramsey spin echo sequence. Qubit 25 is phase shifted by $\pi$ with a pulse starting delay of $10\ \mu$s at a $88(3)\ \mu$m distance from the crystal center (denoted by an $\times$). The dashed green lines in the plot show the population contrast for ion 25 given the SPAM and Ramsey sequence errors, but without the addressing operation. The orange squares show the global microwave pulses, which rotate all qubits by an angle of $\pi/2$ or $\pi$ on the Bloch sphere. The green block represents the addressing ACSS pulse. The inset shows a derotated camera image of the crystal used for this data with a $50\ \mu m$ scale bar.}
    \label{fig:pulsedaddressinghighradius}
\end{figure}

 In order to demonstrate extension of the addressing technique to larger ensembles, Figures \ref{fig:pulsedaddressing} and \ref{fig:pulsedaddressinghighradius} show individual ion addressing data in larger triangular planar crystals ($> 100$ ions).
 
 Not all ions are visible in the associated camera images due to the finite size of the detection laser beam \footnote{The detection and cooling beam waists were limited due to the available laser power. We estimate $>100$ are present in both crystals.}. In the experiment of Fig. \ref{fig:pulsedaddressing}, the addressing laser beam is focused at the radius of the first six ions surrounding the center ion. The starting delay of a $5\ \mu s$-long pulse is varied over a rotation period. Because the pulse length is much shorter than the interval between ion transits, at most one ion receives a $\pi$ phase shift for any given starting delay and has its final population inverted. The ion index spirals outward starting from the ion in the center of the crystal.

Figure \ref{fig:pulsedaddressinghighradius} shows addressing of ion number $25$ at a radius of $88\ \mu$m using a $3\ \mu$s-long pulse ($23$ kHz rotation frequency). At the optimal starting delay of $10.5\ \mu$s, this ion receives a $\pi$ phase shift, while the neighboring ions, numbers 24 and 26, are not shifted. At other nearby starting delays, there is crosstalk to the neighboring ions as they partially transit the beam. The dashed lines for each ion in Fig. \ref{fig:pulsedaddressinghighradius} correspond to the measured bounds for the bare Ramsey spin echo contrast without any addressing operations. This contrast is limited primarily by decoherence from the trap's magnetic field inhomogeneities as explained below. 

Due to the finite thermal occupation of the motional modes and the magnetic field inhomogeneity, each experimental realization results in a different accumulated qubit phase (because of varying Zeeman shifts). Further shimming of the magnetic field or construction of higher-homogeneity magnets would minimize this effect. Also, in the limit of sufficiently low temperatures, achievable through sub-Doppler cooling, the ions will sample the same calibrated magnetic field over many experiments. 

Sub-Doppler cooling has been demonstrated in other Penning traps \cite{jordan_near_2019,stutter_sideband_2018}. Electromagnetically induced transparency (EIT) cooling can efficiently cool the axial modes of motion of hundreds of ions \cite{morigi_cooling_2003,jordan_near_2019}. The radial modes could possibly be cooled in turn via coherent energy exchange between radial and axial modes. This would bring the system near to the motional ground state with mode thermal extents of $<100$ nm, which will also allow for high-fidelity global entangling operations using the metastable qubit demonstrated here.

In order to address every ion in future experiments rather than those at equal radius, an acousto-optic or electro-optic deflector could steer the addressing beam. Due to the rotation, only one-dimensional translation along the crystal radius is sufficient to address all ions. At large radii, the increased ion velocity limits the addressing interaction time, so that greater laser intensity is required. The achievable laser intensity therefore sets an upper bound on the addressable ion radius assuming a fixed phase shift per transit. For $\pi$ phase shift, +1 THz laser detuning, 100 mW in a 10 $\mu$m beam waist, and a $22$ kHz rotation frequency, the maximal addressing radius is $> 2.4$ mm ($>1000$ ions). Alternatively, an ultrafast laser and pulse picker would provide similar addressed phase shifts in shorter time ($<<70$ ns). This would enable addressing in crystals containing thousands of ions.

In conclusion, we have demonstrated individually-addressed single-qubit operations in rotating planar crystals in a compact Penning trap. This implementation utilizes a focused ACSS laser beam to provide single-qubit $\sigma_z$ operations with global microwave $\sigma_x$ and $\sigma_y$ operations. Individual readout is performed efficiently in the rotating frame using an ultra-fast, position-sensitive camera. The operations were performed using a metastable D$_{5/2}$ qubit encoding in \Ca. This encoding leverages mature photonics in the near-IR. With the inclusion of previously demonstrated global entangling operations \cite{bohnet_quantum_2016}, individual addressing and single ion readout form the necessary set of operations to perform quantum simulation, such as in the Quantum Approximate Optimization Algorithm \cite{farhi_quantum_2014, rajakumar_generating_2022}. More generally, the addition of addressing to the Penning trap platform expands the accessible range of quantum many-body simulations with large spin ensembles. 

\begin{acknowledgements}
The authors thank John Bollinger and Bryce Bullock for helpful discussions. 
This material is based upon work supported by the Defense Advanced Research Projects Agency (DARPA) under Contract No. HR001120C0046.
\end{acknowledgements}

\bibliographystyle{apsrev4-2}
\bibliography{refs}

\end{document}